\documentstyle[12pt,epsf,fullpage]{article}

\begin{document}

\title{Finite time singularities in a class of hydrodynamic models}
\author {V.~P. Ruban$^{1,2}$\footnote{Electronic address: ruban@itp.ac.ru},
~D.~I. Podolsky$^1$, ~and ~J.~J. Rasmussen$^2$\\
{\footnotesize $^{1}${\it L.D. Landau Institute for Theoretical Physics,
2 Kosygin Street, 117334 Moscow, Russia}}\\
{\footnotesize $^2${\it Optics and Fluid Dynamics Department, OFD-128,}}
{\footnotesize {\it Ris{\o} National Laboratory, DK-4000 Roskilde, Denmark}}}

\maketitle
\begin{abstract}
Models of inviscid incompressible fluid are considered, with the
kinetic energy (i.e., the Lagrangian functional) taking the form
${\cal L}\sim\int k^\alpha|{\bf v_k}|^2d^3{\bf k}$ in 3D Fourier
representation, where $\alpha$ is a constant, $0<\alpha< 1$.
Unlike the case  $\alpha=0$ (the usual Eulerian hydrodynamics), a
finite value of $\alpha$ results in a finite energy for a
singular, frozen-in vortex filament. This property allows us to
study the dynamics of such filaments without the necessity of a
regularization procedure for short length scales. 
The linear analysis of small
symmetrical deviations from a stationary solution is performed
for a pair of anti-parallel vortex filaments and an analog of the
Crow instability is found at small wave-numbers. A local
approximate Hamiltonian is obtained for the nonlinear long-scale
dynamics of this system. Self-similar solutions of the
corresponding equations are found analytically. They describe the
formation of a finite time singularity, with all length scales
decreasing like $(t^*-t)^{1/(2-\alpha)}$, where $t^*$ is the
singularity time.

\bigskip


\end{abstract}

\section{Introduction}

The question of the possibility for the spontaneous formation of a
finite time singularity  in solutions of the Euler equation for an
ideal incompressible fluid has been discussed for a long time.
At present,  this fundamental problem of fluid
dynamics is still far from a complete solution, though some
rigorous analytical results have been obtained. 
So, it has been found by Beale, Kato and Majda \cite{BKM1984}, 
that no singularity occurs if the time-integral
of the maximum of the vorticity magnitude is finite.
%$$
%\int_{0}^{t^*} \max_{\bf r}|{\bf \Omega}({\bf r},t)|dt <\infty.
%$$
Another result (Constantin and co-workers, \cite{CF1993}-\cite{CFM1996}) 
postulates that a blow-up of the vorticity, if does take place, 
must be accompanied by a singularity in the field of the vorticity direction.
In general, the nature of the presumed
singularity has not yet been clarified, although many theoretical
scenarios for blow-up have been suggested until now, and also
extensive numerical simulations have been performed to observe
the singular behavior (see \cite{PS}-\cite{Moffatt2000} and
references therein). In particular, the locally self-similar 
regime of singularity formation seems very probable. For this
regime, a region of finite size may be distinguished in flow, 
where all length scales, corresponding to vorticity distribution,
decrease like $(t^*-t)^{1/2}$, the velocity increases according 
to $(t^*-t)^{-1/2}$, and the maximum of the vorticity behaves like 
$(t^*-t)^{-1}$. It is necessary to emphasize that this is the only 
possible scaling, which is compatible both with the dimensional 
structure of the Euler equation and with the freezing-in property of 
the vorticity. In this process, an accelerated straining of vortex lines
takes place, and it is the reason for amplification of the vorticity 
magnitude. It is very important that the curvature of vortex lines in the 
assumed self-similar solutions should tend to infinity in the vicinity 
of the singular point, in accordance with the result of Constantin, 
Fefferman and Majda. Thus, the problem of singularity in the 
curvature of vortex lines and the problem of singularity in their 
local stretching are closely connected.

In this paper, we take the point of view that infinite curvature
of frozen-in vortex lines is in some sense a more fundamental
characteristics of hydrodynamic singularity than infinite value
of the vorticity maximum. To illustrate this statement, we
consider a class of models of an incompressible inviscid fluid,
different from Eulerian hydrodynamics, such that finite energy
solutions with infinitely thin frozen-in vortex filaments of
finite strengths are possible. Thus, we deal with a situation  
when the vorticity maximum is infinite from the very beginning,
but nevertheless, this fact itself does not imply a singular
behaviour in the dynamics of vortex strings, while their shape is
smooth and the distance between them is finite. However, the
interaction between filaments may result in formation of a finite
time singularity for the curvature of vortex strings. It is the
main purpose of present work to study this phenomenon analytically.

It is a well known fact that absence  of solutions with singular
vortex filaments in Eulerian hydrodynamics is manifested, in
particular, as a logarithmic divergency of the corresponding formal
expression for the energy functional of an infinitely thin vortex
filament having a finite circulation $\Gamma$ and a shape ${\bf
R}(\xi)$ (this is actually the Hamiltonian functional determining
entirely the dynamics of the system, as is shown in Section 2):
\begin{equation}\label{diverg}
{\cal H}^\Gamma\{{\bf R}(\xi)\}=\frac{\Gamma^2}{8\pi}\oint\!\oint\frac
{({\bf R}'(\xi_1)\cdot{\bf R}'(\xi_2))d\xi_1\,d\xi_2}
{|{\bf R}(\xi_1)-{\bf R}(\xi_2)|}\to\infty.
\end{equation}
More important is that the self-induced velocity of a curved
string in Eulerian hydrodynamics is also infinite. This is the
reason, why we cannot work in the framework of Eulerian
hydrodynamics with such one-dimensional objects, that are very
attractive for theoretical treatment. The situation becomes more
favourable, when we consider a class of regularized models, with
the divergency of the energy functional eliminated. It should be
stressed here that in regularized systems the usual relation
${\bf \Omega}=\mbox{curl }{\bf v}$ between the vorticity and
velocity fields is no more valid, and in this case $\Gamma$ is
not the circulation of the velocity around the filament, but it
is the circulation of the canonical momentum field (see 
Section 2 for more details). However, dynamical properties of a
de-singularized system depend on the manner of regularization.
For instance, it is possible to replace the singular Green's
function  $G(|{\bf R}_1-{\bf R}_2|)$ in (\ref{diverg}) 
(where $G(r)\sim 1/r$) by some analytical function which has no 
singular points near the real axis in the complex plane 
(for examples by $G_q(r)\sim\tanh (qr)/r$ or by
$G_\epsilon(r)\sim1/\sqrt{r^2+\epsilon^2}$). In that case we may
not expect any finite time singularity formation, because the 
corresponding velocity field created by the vortex string appears 
to be too smooth with any shape of the curve, and this fact prevents 
drawing together some pieces of the string. With such a very smooth
velocity field, a singularity formation needs an infinite time.

In this paper we consider another type of regularization of the
Hamiltonian functional, when the Green's function is still
singular, but this singularity is integrable in the contour
integral analogous to the expression (\ref{diverg}):
\begin{equation}\label{H_alpha}
{\cal H}_\alpha^\Gamma\{{\bf R}(\xi)\}\sim
\frac{\Gamma^2}{2}\oint\!\oint\frac
{({\bf R}'(\xi_1)\cdot{\bf R}'(\xi_2))d\xi_1\,d\xi_2}
{|{\bf R}(\xi_1)-{\bf R}(\xi_2)|^{1-\alpha}},
\end{equation}
with a small but finite positive constant $0<\alpha\ll 1$. If
$\alpha$ is not small,  we actually have models that are rather
different from Eulerian hydrodynamics. Nevertheless, such models
still have many common features with usual hydrodynamics, which
are important for singularity formation in the process of the
interaction between  vortex filaments: a similar hydrodynamic
type structure of the Hamiltonian and a power-like behaviour of
the Green's function, with negative exponent. Therefore we
believe that it is useful to investigate these models, especially
the question about the formation of a finite time singularity  in
the vortex line curvature. We hope the results of our study will
shed more light on the problem of blow-up in Eulerian
hydrodynamics.

This paper is organized as follows. In  Section II, we briefly
review some basic properties of frozen-in vorticity dynamics in a
perfect fluid, with giving necessary definitions for theoretical
conceptions used in our study. In general, our approach is based
on the Hamiltonian formalism for frozen-in vortex lines
\cite{Berdichevsky}-\cite{R2000PRE}. Then, in  Section III, we
perform the linear analysis of stability for a pair of symmetric
anti-parallel vortex filaments and find an instability at small
wave numbers analogous to the Crow instability \cite{Crow}. In
Section IV, we postulate a local approximate Hamiltonian for the
long scale nonlinear dynamics of the pair of filaments and
present analytical self-similar solutions of the corresponding
equations. Those solutions describe finite time singularity
formation, with the length scales decreasing like
$(t^*-t)^{1/(2-\alpha)}$, and this is the main result of present
work. In Section V, we make some concluding remarks about vortex
filaments of a finite width, about long scale approximation for
systems with the Green's function of a general form, and finally
about how it is possible to improve the approximation in the case
of small $\alpha$, when the unstable region is narrow in wave
number space. In Appendix A, we write in terms of the special
mathematical functions some integral expressions needed for
calculation of instability increment of the vortex pair. In
Appendix B, we provide details about the integration procedure
for the system of ordinary differential equations related to the
self-similar solutions.

\section{Hamiltonian dynamics of vortex filaments}

To clarify the meaning of the suggested models (\ref{H_alpha}) and
to explain the employed theoretical method, we recall some
general properties of frozen-in vorticity dynamics in a perfect
fluid, starting from the Lagrangian formalism
\cite{Arnold}-\cite{IL}, \cite{Berdichevsky}-\cite{R2000PRE}.

Let a Lagrangian functional ${\cal L}\{{\bf v}\}$ specify the
dynamics of some incompressible medium of unit density, with the
solenoidal velocity field ${\bf v}({\bf r},t)$. Especially we are
interested here in systems with quadratic Lagrangians, which in
3D Fourier representation take the form:
\begin{equation}\label{L_M}
{\cal L}_M\{{\bf v}\}=\frac{1}{2}\int\frac{d^3{\bf k}}{(2\pi)^3}
M(k)|{\bf v}_{\bf k}|^2,
\end{equation}
where $M(k)$ is some given positive function of the absolute
value of the wave vector ${\bf k}$. This expression should be
understood as the kinetic energy on the group of volume
preserving mappings ${\bf x}({\bf a},t)$, and the velocity field
${\bf v}({\bf x},t)$ is defined as the time derivative $\dot {\bf
x}({\bf a},t)$ taken at the point ${\bf a}({\bf x},t)$. Obviously,
all the systems (\ref{L_M}) possess the properties of homogeneity and
isothropy in the space. It is
clear that the usual Eulerian hydrodynamics corresponds to the
simplest case $M(k)=1$. Another physically important example concerns
the homogeneous incompressible electron magnetohydrodynamics (EMHD), 
for which $M(k)=1+q^2/k^2$, with a constant $q$ beeng the screening 
parameter \cite{R2000PRE},\cite{physics/0007010}. 
Also the case $M(k)=1+\lambda^2k^2$ has been studied, 
with a constant $\lambda$, which corresponds to the so called 
avereged Eulerian hydrodynamics (see, for instance, the papers \cite{HMR98}, 
\cite{HKMRS99} for more details). In the general case, the systems 
(\ref{L_M}) may be understood as models for some inviscid non-newtonian 
fluids. It should be noted that there exists a direct relation between such 
models and the vortex blob method introduced by Chorin for 
desingularization of the Eulerian hydrodynamics \cite{Chorin73}. 
Some discussion of this relation, for the case of the averaged Eulerian
hydrodynamics, can be found in the papers \cite{OS99} and \cite{NS2000}. 

Due to the presence of the Noether type symmetry with respect to relabeling
of Lagrangian labels of fluid points \cite{Salmon}-\cite{IL},
\cite{Berdichevsky}-\cite{R99}, all such systems have an infinite number of
integrals of motion, which can be expressed as conservation of the circulations
$\Gamma_c$ of the canonical momentum field ${\bf p}({\bf r},t)$,
\begin{equation}
{\bf p}=\frac{\delta{\cal L}}{\delta{\bf v}},
\end{equation}
along any closed contour $c(t)$ advected by flow, thus the 
generalized theorem of Kelvin is valid:
\begin{equation}
\Gamma_c=\oint_{c(t)} ({\bf p}\cdot d{\bf l})=\mbox{const}.
\end{equation}
These integrals of motion correspond to the frozen-in property of
the canonical vorticity field  ${\bf \Omega}({\bf r},t)$,
\begin{equation}\label{Omega_definition}
{\bf \Omega}\equiv\mbox{curl }{\bf p}=
\mbox{curl }\frac{\delta{\cal L}}{\delta{\bf v}}.
\end{equation}
After defining the Hamiltonian functional ${\cal H}\{{\bf \Omega}\}$,
\begin{equation}\label{Ham_definition}
{\cal H}\{{\bf \Omega}\}=\Big(\int \Big(\
\frac{\delta{\cal L}}{\delta{\bf v}}\cdot{\bf v}\Big)d{\bf r} \,\,
-{\cal L}\Big)\Big|_{{\bf v}={\bf v}\{{\bf \Omega}\}},
\end{equation}
the equation of motion for the vorticity takes the form
\begin{equation}\label{Ham}
{\bf\Omega}_t=\mbox{curl}
\left[\mbox{curl}\left(\frac{\delta{\cal H}}{\delta{\bf\Omega}}\right)
\times{\bf\Omega}
\right].
\end{equation}
This equation describes the transport of frozen-in vortex lines
by the flow having the velocity field
\begin{equation}
{\bf v}=\mbox{curl}\left(\frac{\delta{\cal H}}{\delta{\bf\Omega}}\right).
\end{equation}
It is very important  in this process that all topological
characteristics of the vorticity field are conserved
\cite{Arnold}, \cite{MonSas}, \cite{RA94}.
It follows from Eqs. (\ref{L_M}), (\ref{Omega_definition}) and 
(\ref{Ham_definition}), that the Hamiltonian ${\cal H}_M$ 
corresponding to the Lagrangian ${\cal L}_M$ is
\begin{equation}\label{H_M}
{\cal H}_M\{{\bf \Omega}\}=\frac{1}{2}\int\frac{d^3{\bf k}}{(2\pi)^3}
\frac{|{\bf \Omega}_{\bf k}|^2}{k^2M(k)}
=\frac{1}{2}\int\!\!\int
G_M(|{\bf r}_1-{\bf r}_2|)({\bf \Omega}({\bf r}_1)\cdot{\bf \Omega}({\bf r}_2)
d{\bf r}_1d{\bf r}_2,
\end{equation}
with the Green's function $G_M(r)$ being equal to the following integral:
\begin{equation}\label{G_M}
G_M(r)=\int\frac{d^3{\bf k}}{(2\pi)^3}\frac{e^{i{\bf kr}}}{k^2M(k)}
=\frac{1}{2\pi^2}\int\limits_0^{+\infty}\frac{\sin kr}{kr}\frac{dk}{M(k)}.
\end{equation}

The frozen-in  vorticity field can be represented in topologically simple
cases as a continuous distribution of vortex lines
\cite{Berdichevsky}-\cite{R2000PRE}:
\begin{equation}\label{lines}
{\bf \Omega }({\bf r},t)=\int_{\cal N}d^2\nu \oint \delta ({\bf r}-
{\bf R}(\nu,\xi,t))\frac{\partial{\bf R}}{\partial\xi}d\xi,
\end{equation}
where a 2D Lagrangian coordinate $\nu=(\nu_1,\nu_2)$, which lies
in some manifold ${\cal N}$, is the label of a vortex line, while
the longitudinal coordinate $\xi$ determines a point on the line.

The important characteristics of the system: the (virtual) linear momentum 
${\bf P}$ and the angular momentum ${\bf M}$ can be expressed as
follows:
\begin{eqnarray}
{\bf P}&=&\int_{\cal N}d^2\nu\,\,\frac{1}{2}\oint[{\bf R}\times{\bf R}_\xi]d\xi,
\label{mom}\\
{\bf M}&=&
\int_{\cal N}d^2\nu\,\,\frac {1}{3}\oint {\bf [R\times [R\times R}_\xi]]d\xi.
\label{anglmom}
\end{eqnarray}

In the limit, when the shapes ${\bf R}(\nu,\xi,t)$ of vortex lines do not
depend on the label $\nu$, we have one singular vortex filament with a finite
circulation $\Gamma=\int_{\cal N}d^2\nu$. In this case, the flow is
potential in the space around the filament: ${\bf p}=\nabla\Phi$, with a
multi-valued scalar potential $\Phi({\bf r},t)$. The potential flow domain is
passive from the dynamical viewpoint, because there the flow depends entirely
on the filament shape. The dynamics of the shape ${\bf R}(\xi,t)$ of
such infinitely thin vortex filament is determined in a self-consistent manner
by the variational principle with the Lagrangian
${\cal L}_M^{\Gamma}\{{\bf R}\}$ \cite{Berdichevsky}-\cite{R2000PRE},
\begin{equation}\label{LGamma_M}
{\cal L}_M^{\Gamma}=\Gamma\oint([{\bf R}'\times{\bf R}_t]\cdot
{\bf D}({\bf R}))d\xi
-\frac{\Gamma^2}{2}\oint\!\!\oint
G_M\left(|{\bf R}(\xi_1)-{\bf R}(\xi_2)|\right)\,
\left({\bf R}'(\xi_1)\cdot{\bf R}'(\xi_2)\right)\,d\xi_1d\xi_2,
\end{equation}
where the vector function ${\bf D}({\bf R})$ must have unit divergence
\cite{R2000PRE}:
\begin{equation}
\mbox{div}_{\bf R}{\bf D}({\bf R})=1.
\end{equation}
The generalization of the expression (\ref{LGamma_M}) to a case of several
filaments with the circulations $\Gamma^{(n)}$ and shapes
${\bf R}^{(n)}(\xi,t)$, $n=1..N$, is straightforward: one should write
a single sum over $n$ for the first term and a double sum for the Hamiltonian.

It is easy to see that the Hamiltonian (\ref{H_alpha}) corresponds to the
function $M(k)$ in the form
\begin{equation}
M(k)\sim k^\alpha.
\end{equation}

The choice of the longitudinal parameter $\xi$ is not unique, but
this does not affect the dynamics of the vortex string which is
an invariant geometric object. Sometimes it is convenient to use
parameterization of the vortex line shape by the Cartesian
coordinate:
\begin{equation}\label{Cartesian}
{\bf R}(\xi,t)=(X(\xi,t),Y(\xi,t),\xi).
\end{equation}
Then the choice ${\bf D}=(0,Y,0)$ gives immediately that $X(\xi,t)$ and
$Y(\xi,t)$ are canonically conjugated quantities.

Hereafter, we will consider vortex filaments with unit circulation for
simplicity. So the symbol $\Gamma$, if appearing in some expressions below,
will mean the special mathematical Gamma function. Also, without loss of 
generality, all quantities may be considered as dimensionless.

Now, for some fixed value of the parameter $\alpha$, let us
consider the symmetrical dynamics of a pair of oppositely
rotating vortex filaments, with the symmetry plane $y=const$. Due
to this symmetry, it is sufficient to consider only one of the
filaments. It follows from the above discussion that the exact
expression for the Hamiltonian of this system is the following:
$$
{\cal H}_\alpha=\frac{1}{2}\int\!\!\int\!\!
\frac{(1+X_1' X_2'+Y_1' Y_2')\,\,d\xi_1 \,d\xi_2}
{\Big((\xi_1-\xi_2)^2+\!(X_1-X_2)^2+\!(Y_1-Y_2)^2\Big)^{\frac{1-\alpha}{2}}}
$$
\begin{equation}\label{Hamiltonian}
+\frac{1}{2}\int\!\!\int\!\frac{(-1-X_1' X_2'+Y_1' Y_2')\,\,d\xi_1 \,d\xi_2}
{\Big((\xi_1-\xi_2)^2+\!(X_1-X_2)^2+\!(Y_1+Y_2+b)^2\Big)^{\frac{1-\alpha}{2}}},
\end{equation}
where $b$ is the mean distance between the two filaments ($b$
does not depend on time because of the conservation law for the
momentum (\ref{mom})), $X_1=X(\xi_1)$, $X_1'=X'(\xi_1)$ and so
on. The first term in Eq.(\ref{Hamiltonian}) describes the
non-local self-interaction of the filament, while the second one
corresponds to the interaction with the second filament. The
Hamiltonian equations of motion have the form
\begin{equation}\label{Ham_eqs}
\dot X(\xi)=\frac{\delta{\cal H}_\alpha}{\delta Y(\xi)}, \qquad
\dot Y(\xi)=-\frac{\delta{\cal H}_\alpha}{\delta X(\xi)}.
\end{equation}

\section{Crow instability for a pair of vortex filaments}

The system with the Hamiltonian (\ref{Hamiltonian}) possesses the exact
stationary solution
\begin{equation}
X(\xi,t)=C(\alpha,b)t,\qquad Y(\xi,t)=0,
\end{equation}
which describes the uniform motion of straight filaments. Here the stationary
velocity $C(\alpha,b)$ is proportional to $b^{\alpha-1}$. But this solution
appears to be unstable due to an analog of the Crow instability \cite{Crow}.
In this section we consider the linear evolution of small perturbations of the
stationary solution, and derive the linear growth rate.

To perform the linear analysis of small deviations of the vortex shape from a
straight line, we need the quadratic part of the Hamiltonian
(\ref{Hamiltonian}):
$$
{\cal H}^{(2)}_\alpha=\frac{1}{2}\int\!\!\int\frac
{(X_1' X_2'+Y_1' Y_2')}{|\xi_1-\xi_2|^{1-\alpha}}\,d\xi_1 \,d\xi_2
+\frac{1}{2}\int\!\!\int\left(\frac{\alpha-1}{2}\right)
\frac{[(X_1-X_2)^2+(Y_1-Y_2)^2]}
{|\xi_1-\xi_2|^{3-\alpha}}\,d\xi_1 \,d\xi_2
$$
$$
+\frac{1}{2}\int\!\!\int\frac{(Y_1' Y_2'-X_1' X_2')}
{\Big((\xi_1-\xi_2)^2+b^2\Big)^{\frac{1-\alpha}{2}}}\,d\xi_1 \,d\xi_2
-\frac{1}{2}\int\!\!\int\left(\frac{\alpha-1}{2}\right)
\frac{[(X_1-X_2)^2+(Y_1+Y_2)^2]}
{\Big((\xi_1-\xi_2)^2+b^2\Big)^{\frac{3-\alpha}{2}}}\,d\xi_1 \,d\xi_2
$$
\begin{equation}\label{Hquadratic}
-\frac{1}{2}\int\!\!\int\left(\frac{\alpha\!-\!1}{2}\right)
\left(\frac{\alpha\!-\!3}{2}\right)
\frac{2b^2(Y_1+Y_2)^2}
{\Big((\xi_1-\xi_2)^2+b^2\Big)^{\frac{5-\alpha}{2}}}d\xi_1 d\xi_2.
\end{equation}
For further consideration, it is useful to rewrite it in
the 1D Fourier representation:
\begin{equation}\label{FourierHquadratic}
{\cal H}^{(2)}_\alpha=
\frac{1}{2}\int\frac{d k}{2\pi}\Big(A_\alpha(k)X_k X_{-k}
+B_\alpha(k)Y_k Y_{-k}\Big).
\end{equation}

Expressions for the functions $A_\alpha(k)$ and $B_\alpha(k)$
follow from Eq.(\ref{Hquadratic}). So, $A_\alpha(k)$ can be
represented as follows:
$$
A_\alpha(k)=2k^2b^\alpha\int_0^{+\infty}\!\!\cos(kb\zeta)
\left(\frac{1}{\zeta^{1-\alpha}}-\frac{1}
{(\zeta^2+1)^{\frac{1-\alpha}{2}}}\right)d\zeta
$$
$$
+2(\alpha\!-\!1)b^{\alpha-2}\!\!\int_0^{+\infty}\!\!\!(1-\cos(kb\zeta))
\left(\frac{1}{\zeta^{3-\alpha}}\!-
\!\frac{1}{(\zeta^2+1)^{\frac{3-\alpha}{2}}}\right)d\zeta
$$
\begin{equation}\label{Ak}
=2(1-\alpha)^2b^{\alpha-2}\int_0^{+\infty}\!\!(1-\cos(kb\zeta))
\left(\frac{1}{\zeta^{3-\alpha}}-
\frac{1}{(\zeta^2+1)^{\frac{3-\alpha}{2}}}
+\left(\frac{3\!-\!\alpha}{1\!-\!\alpha}\right)
\frac{1}{(\zeta^2+1)^{\frac{5-\alpha}{2}}}\right)d\zeta
\end{equation}
Obviously, $A_\alpha(k)$ is positive everywhere.
Analogous calculations for the function $B_\alpha(k)$ give:
$$
B_\alpha(k)=2k^2b^\alpha\int_0^{+\infty}\!\!\cos(kb\zeta)
\left(\frac{1}{\zeta^{1-\alpha}}+
\frac{1}{(\zeta^2+1)^{\frac{1-\alpha}{2}}}\right)d\zeta
+2(\alpha-1)b^{\alpha-2}\int_0^{+\infty}\!\!(1-\cos(kb\zeta))
\frac{d\zeta}{\zeta^{3-\alpha}}
$$
$$
+2(1-\alpha)b^{\alpha-2}\int_0^{+\infty}\!\!(1+\cos(kb\zeta))
\left(\frac{1}{(\zeta^2+1)^{\frac{3-\alpha}{2}}}
+\frac{\alpha-3}{(\zeta^2+1)^{\frac{5-\alpha}{2}}}\right)d\zeta
$$
$$
=2(1-\alpha)^2b^{\alpha-2}\int_0^{+\infty}\!\!(1-\cos(kb\zeta))
\frac{d\zeta}{\zeta^{3-\alpha}}
$$
\begin{equation}\label{Bk}
-2(1-\alpha)(3-\alpha)b^{\alpha-2}\int_0^{+\infty}\!\!(1+\cos(kb\zeta))
\left(\frac{2}{(\zeta^2+1)^{\frac{5-\alpha}{2}}}
-\frac{1}{(\zeta^2+1)^{\frac{3-\alpha}{2}}}\right)d\zeta
\end{equation}
In Appendix A, $A_\alpha(k)$ and $B_\alpha(k)$ are expressed through the Euler
Gamma function $\Gamma(x)$ and the modified Bessel functions of the second
kind $K_\nu(x)$.

\begin{figure}
\epsfxsize=4.5in
\epsfysize=2.5in
\centerline{\epsffile{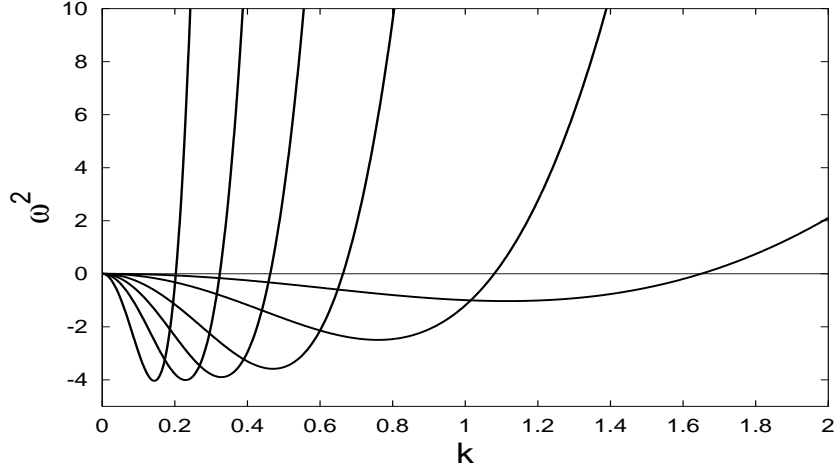}}
\caption{\small The dependences $\omega_\alpha^2(k)=A_\alpha(k)B_\alpha(k)$
with $b=1$ for $\alpha=0.01$, $0.025$, $0.05$, $0.1$, $0.25$, $0.5$.
Lines corresponding to the given values of $\alpha$ intersect the horizontal
axis in the indicated order.}
\label{AB}
\end{figure}

The dispersion relation between the frequency $\omega_\alpha$ of a small amplitude
perturbation of the filament shape and the corresponding wave number $k$ is
simply given by the formula
\begin{equation}\label{omega_k}
\omega_\alpha^2(k)=A_\alpha(k)B_\alpha(k),
\end{equation}
since the linearized equations of motion for $X_k$ and $Y_k$  are
\begin{equation}\label{dotXkYk}
\dot X_k=B_\alpha(k)Y_k,\qquad \dot Y_k=-A_\alpha(k)X_k,
\end{equation}
as follows from Eq.(\ref{Ham_eqs}). In Fig.\ref{AB} we have plotted
$\omega_\alpha^2$ versus $k$ for several values of $\alpha$.

\begin{figure}
\epsfxsize=4.5in
\epsfysize=2.5in
\centerline{\epsffile{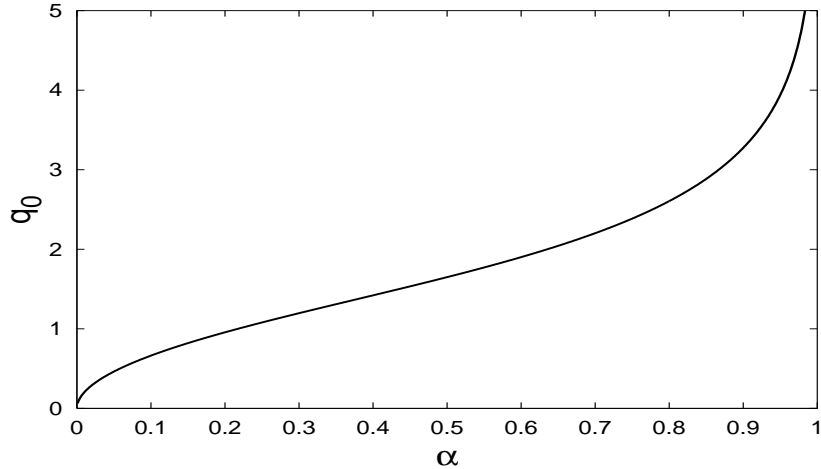}}
\caption{\small The boundary of instability $q_0(\alpha)$.}
\label{q_0_alpha}
\end{figure}

It is easy to see that at small wave numbers the product
$A_\alpha(k)B_\alpha(k)$ is negative. Indeed, after some calculations
we obtain in leading order for $kb\ll 1$:
\begin{equation}\label{Alongscale}
A_\alpha(k)\approx k^2b^\alpha\left(\frac{1-\alpha}{\alpha}\right)I_{3-\alpha},
\end{equation}
\begin{equation}\label{Blongscale}
B_\alpha(k)\approx -4(1-\alpha)^2 b^{\alpha-2}I_{3-\alpha},
\end{equation}
where the constant $I_{3-\alpha}$ is given by the integral
\begin{equation}
I_{3-\alpha}=\int\limits_0^{+\infty}
\frac{d\zeta}{(\zeta^2+1)^{\frac{3-\alpha}{2}}}
=\frac{\sqrt{\pi}\Gamma\left(1-\frac{\alpha}{2}\right)}
{2\Gamma\left(\frac{3-\alpha}{2}\right)},
\end{equation}
with $\Gamma(..)$ being the Gamma function. Therefore, an
instability takes place at small $k$. The unstable domain in the
wave number space corresponds to a range $|k|b<q_0(\alpha)$ where
$B_\alpha(k)$ is negative, with the function $q_0(\alpha)$
behaving, at small values of $\alpha$, like $\sqrt{\alpha}$:
\begin{equation}\label{q_0}
q_0(\alpha)\approx 2\sqrt{\alpha}, \qquad \alpha\ll 1.
\end{equation}
The plot of $q_0(\alpha)$ is shown in Fig.\ref{q_0_alpha}.
The instability increment $\gamma_\alpha(k)=\sqrt{-A_\alpha(k)B_\alpha(k)}$
is proportional to the absolute value of $k$ at very small values of $kb$:
\begin{equation}\label{increment}
\gamma(k)\approx (1-\alpha)I_{3-\alpha}\cdot 2|k|b^{\alpha-1}
\sqrt{({1-\alpha})/{\alpha}}.
\end{equation}
However, for each $\alpha$ there exists a maximum value
$\gamma_{\mbox{\scriptsize max}}(\alpha)$ of the increment, which
is attained at $k b\sim \sqrt{\alpha}$. Therefore the approximate
expressions (\ref{Blongscale}) and (\ref{increment}) are valid
only if $|k|b\ll \sqrt{\alpha}$. 
%It is interesting to note that the following inequality takes place:
%$\gamma_{\mbox{\scriptsize max}}(\alpha)<2b^{\alpha-2}$
%(see the Fig.\ref{AB}, where for the case $b=1$ the minimal value 
%of the product $A_\alpha(k)B_\alpha(k)$ approaches  $-4$ as $\alpha\to 0$). 

For large wave numbers, $|k|b\gg 1$, the functions $A_\alpha(k)$ and
$B_\alpha(k)$ are both positive. The asymptotic approximations in that
region are:
\begin{equation}\label{short_scale}
A_\alpha(k)\approx B_\alpha(k)\approx 2(1-\alpha)^2k^{2-\alpha}\int_0^{+\infty}
\frac{(1-\cos \eta)}{\eta^{3-\alpha}}d\eta
=k^{2-\alpha}\frac{2(1-\alpha)\cos(\pi\alpha/2)\Gamma(\alpha)}{2-\alpha}.
\end{equation}
Note that this expression does not contain the parameter $b$.
For a single vortex filament it is actually the exact expression for
$A_\alpha(k)$ and $B_\alpha(k)$, which is valid in the whole range of $k$.

A general nonlinear analysis of the non-local system (\ref{Hamiltonian})
is difficult. Therefore we need some simplified model which would approximate
the nonlinear dynamics, at least in the most interesting long scale unstable
regime. In the next section, we suggest such an approximate model and find
a class of solutions describing the formation of a finite time singularity.

\section{Singularity in long-scale nonlinear dynamics}

We note that the same long-scale limit as
(\ref{Alongscale}-\ref{Blongscale})
can be obtained from the local nonlinear Hamiltonian
\begin{equation}\label{H_l}
{\cal H}_l\{{\bf R(\xi)}\}=
(1-\alpha)I_{3-\alpha}\oint\frac{(2Y)^\alpha}{\alpha}\sqrt{X'^2+Z'^2}d\xi,
\end{equation}
where the coordinate $Y(\xi)$ is measured from the symmetry plane. This
Hamiltonian approximates the exact non-local Hamiltonian of a symmetrical
pair of vortex filaments in the case when the ratio of a typical value of $Y$
to a typical longitudinal scale $L$ is much smaller than $q_0(\alpha)$:
\begin{equation}\label{local_condition}
Y/L\ll\sqrt{\alpha}.
\end{equation}
In particular, this means that the slope of the curve with
respect to the symmetry plane should be small, and also $Y$
should be small in comparison with the radius of the line
curvature. When $Y=\mbox{const}$, $X'=\mbox{const}$,
$Z'=\mbox{const}$, expression (\ref{H_l}) gives the same result
for uniform stationary motion as the exact Hamiltonian.

With the Cartesian parametrization (\ref{Cartesian}), the corresponding
approximate local nonlinear equations of motion have the form
(after appropriate time rescaling)
\begin{eqnarray}
\dot X &=& \frac{1}{(2-\alpha)}
\frac{\sqrt{1+X'^2}}{Y^{1-\alpha}},\label{dotXz}\\
\dot Y &=& \frac{1}{(2-\alpha)\alpha}\left(
\frac{Y^\alpha X'}{\sqrt{1+X'^2}}\right)' \label{dotYz}
\end{eqnarray}
and they allow to obtain a simple explanation of the instability.
On a qualitative level of understanding, the reason for the instability is
that if initially some pieces of the curve were more close to the symmetry
plane and convex in the direction of motion, then at subsequent moments
in time the curvature will be increasing because of smaller values of $Y$
and corresponding larger velocity, while $Y$ will be decreasing due to the
curvature. Thus, the feedback is positive and the system is unstable.
In the final stage of the instability development, a locally self-similar
regime in the dynamics is possible, because the above equations
admit the self-similar substitution
\begin{eqnarray}
X(\xi,t)&=&X^*-(t^*-t)^\beta x\left((\xi-\xi^*) (t^*-t)^{-\beta}\right),
\label{Xselfsimsubs}\\
Y(\xi,t)&=&(t^*-t)^\beta y\left((\xi-\xi^*) (t^*-t)^{-\beta}\right),
\label{Yselfsimsubs}
\end{eqnarray}
with arbitrary constants $X^*$, $\xi^*$, $t^*$, and with the exponent
\begin{equation}\label{beta}
\beta=\frac{1}{2-\alpha}.
\end{equation}
After substituting Eqs.(\ref{Xselfsimsubs}-\ref{Yselfsimsubs})
into Eqs.(\ref{dotXz}-\ref{dotYz}), we obtain a pair of ordinary
differential equations for the functions $x(z)$ and $y(z)$:
\begin{eqnarray}
x-z \cdot\frac{dx}{dz}&=&\frac{\sqrt{1+(dx/dz)^2}}{y^{1-\alpha}},
\label{Xselfsim}\\
y-z \cdot\frac{dy}{dz}&=&\frac{1}{\alpha}\cdot\frac{d}{dz}\left(
\frac{y^\alpha \cdot(dx/dz)}{\sqrt{1+(dx/dz)^2}}\right)
\label{Yselfsim},
\end{eqnarray}
where $z = (\xi-\xi^*) (t^*-t)^{-\beta}$.

\begin{figure}
\epsfxsize=4.5in
\epsfysize=2.0in
\centerline{\epsffile{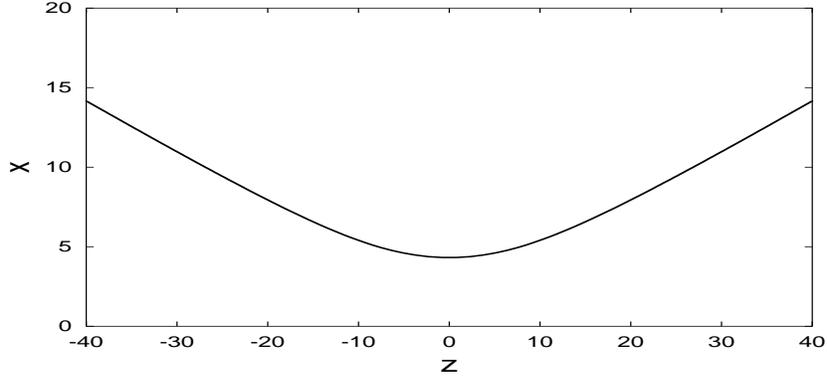}}
\caption{\small Self-similar solution $x(z)$ for $C=50$, $\alpha=0.1$.}
\label{xz}
\end{figure}

However, with this
choice of  parametrization of the curve, the obviously existing
symmetry of the system (\ref{H_l}) with respect to rotation in
the $x$-$z$ plane is hidden. For taking advantage of this
symmetry, cylindrical coordinates are more appropriate, with the
angle coordinate $\varphi$ serving as the longitudinal parameter:
\begin{equation}
(X,Y,Z)=(R(\varphi,t)\cos\varphi, Y(\varphi,t), -R(\varphi,t)\sin\varphi).
\end{equation}
Instead of the equations of motion (\ref{dotXz}-\ref{dotYz}), we obtain
the equivalent system (where a same time rescaling as in
(\ref{dotXz}-\ref{dotYz}) is performed)
\begin{eqnarray}
-(2-\alpha)R\dot R &=& \frac{\sqrt{R^2+R'^2}}{Y^{1-\alpha}},\label{dotR}\\
-(2-\alpha)R\dot Y &=&
\frac{1}{\alpha}\left(\frac{Y^\alpha R'}{\sqrt{R^2+R'^2}}\right)'
-\frac{1}{\alpha}\frac{RY^\alpha}{\sqrt{R^2+R'^2}}.\label{dotY}
\end{eqnarray}
Here $(..)'=\partial_\varphi(..)$. This system follows from the Lagrangian
written in cylindrical coordinates
\begin{equation}
{\cal L}_\varphi\sim\int\left((2-\alpha)\frac{R^2}{2}\dot Y \,\,
-\frac{Y^\alpha}{\alpha}\sqrt{R^2+R'^2}\right)d\varphi.
\end{equation}
The self-similar substitution
\begin{equation}
R(\varphi,t)=(t^*-t)^\beta r(\varphi), \quad
Y(\varphi,t)=(t^*-t)^\beta y(\varphi)
\end{equation}
does not change the meaning of the angle coordinate $\varphi$. It leads us to
the following pair of equations for the functions $r(\varphi)$ and $y(\varphi)$:
\begin{eqnarray}
r^2 &=& \frac{\sqrt{r^2+r'^2}}{y^{1-\alpha}},\label{r_varphi}\\
yr &=& \frac{1}{\alpha}
\left(\frac{y^\alpha r'}{\sqrt{r^2+r'^2}}\right)'
-\frac{1}{\alpha}\frac{ry^\alpha}{\sqrt{r^2+r'^2}}.\label{y_varphi}
\end{eqnarray}
We observe that there is no explicit dependence on $\varphi$ in
these equations. This property helps us to integrate the system.
The general solution can be represented in the following
parametric form (see Appendix B for a detailed derivation):
\begin{equation}\label{varphi_p}
\varphi(p)=\varphi_0+\arctan (p)
-\sqrt{\frac{\alpha(1-\alpha)}{(2-\alpha)(1+\alpha)}}
\cdot\arctan\left(p\sqrt{\frac{\alpha(2-\alpha)}{(1-\alpha^2)}}\right) ,
\end{equation}
\begin{equation}\label{y_p}
y(p)=C^{-\frac{1}{2-\alpha}}\left(
\frac{(1-\alpha^2)}{\alpha(2-\alpha)}+p^2\right)^{\frac{1}{2(2-\alpha)}},
\end{equation}
\begin{equation}\label{r_p}
r(p)=C^{\frac{1-\alpha}{2-\alpha}}\left(
\frac{(1-\alpha^2)}{\alpha(2-\alpha)}+p^2\right)^{\frac{\alpha-1}{2(2-\alpha)}}
\sqrt{1+p^2},
\end{equation}
where the parameter $p$ runs between the limits
$-\infty<p<+\infty$, $C$ and $\varphi_0$ are arbitrary
integration constants . The constant $C$ determines the asymptotic
slope of the curve at large distances from the origin: $y\approx
r/C$ when $r\to\infty$, while the constant $\varphi_0$ reflects
the mentioned symmetry of the system with respect to rotations in
$x$-$z$ plane. The condition (\ref{local_condition}) for
applicability of the local approximation (\ref{H_l}) is satisfied
if $C\sqrt{\alpha}\gg 1$. A typical self-similar solution $x(z)$
is shown in Fig.\ref{xz}.

It is interesting to note that the total angle $\Delta \varphi$ between two
asymptotic directions in $x$-$z$ plane does not depend on the parameter $C$
in the long-scale local approximation used above:
\begin{equation}
\Delta \varphi=
\pi\left(1-\sqrt{\frac{\alpha(1-\alpha)}{(2-\alpha)(1+\alpha)}}\,\,\right).
\end{equation}
At small values of $\alpha$, this angle approaches $\pi$.
Another remark about $\Delta \varphi$ is that the above expression assumes
identical values at $\alpha$ and at $1-\alpha$, so the value
$\tilde\alpha=1/2$ results in the extremum $\Delta \varphi_{min}=2\pi/3$.
For this case, the curve lies on the cone $y=r/C$.

\section{Discussion}

We observed that in the systems (\ref{H_l}) with $0<\alpha<1$,
finite time singularity formation is possible in the self-similar
regime. Inasmuch as the condition (\ref{local_condition}) for
applicability of the approximate Hamiltonian (\ref{H_l}) is
satisfied in a range of the parameter $C$ related to the
self-similar solutions (\ref{varphi_p}-\ref{r_p}), we conclude
that in the systems (\ref{H_alpha}) self-similar collapse of two
symmetrical singular vortex filaments can also take place. The
principal question is whether this is also possible for filaments
having finite width. If yes, then such solutions are analogous to
the assumed  self-similar solutions of the Euler equation.
Though the exponent $\beta$ (\ref{beta}) differs from $1/2$, the
difference is small if $\alpha$ is small. However, an important
difference exists between infinitely thin filaments and filaments
with finite width: inside the latter, longitudinal flows take
place, caused by a twist of the vortex lines constituting the
filament. Those flows keep the width homogeneous along the
filament if a local stretching is not sufficiently fast. This
mechanism acts against singularity formation and, probably, in
some cases it can prevent a singularity at all. [It is worth
mentioning here that for finite width vortex structures in the
Navier-Stokes equation frame, the usual "outcome" result of the Crow
instability is vortex line reconnection \cite{KT94}.] Thus, 
a more-less consistent analysis of the
general situation should take into account, besides the dynamics
of a mean shape of the filament, at least the dynamics of the
width and the conjugated dynamics of the twist. Clearly, we do
not need to consider $\alpha\not= 0$ systems, when we deal with
non-singular vortex filaments. It should be emphasized that an
attempt to take account of finite width of the filament by simply
using regularized Green's functions such as $G_\epsilon(r)\sim
1/\sqrt{r^2+\epsilon^2}$ with a constant $\epsilon$, giving
correct results for long scale limit of the linearized problem,
fails to describe the dynamics in the highly nonlinear regime.

Also, we would like to note that a local approximation analogous 
to (\ref{H_l}) is possible for arbitrary Green's function $G_M(r)$. 
The corresponding long scale Hamiltonian has the form
$$
{\cal H}_{Ml}\{{\bf R}(\xi)\}=\oint F_M(Y)\sqrt{X'^2+Z'^2}d\xi,
$$
where the (positive) function $F_M(Y)$ is related to the function 
$G_M(r)$ in the following way:
$$
F_M(Y)=\int_0^{+\infty}\left(G_M(\xi)-G_M\Big(\sqrt{\xi^2+(2Y)^2}\Big)
\right)d\xi.
$$
The stationary motion with a constant coordinate $Y_0=b/2$ is unstable if the
second derivative of the function $F_M$ is negative at that value:
$F_M''(b/2)<0$. We believe that such systems can exhibit locally self-similar
collapse, if the asymptotics of the function $F_M(Y)$ is
power-like at small $Y$: $F_M\sim Y^\alpha$, with $0<\alpha<1$.

The final remark concerns the possibility of including  effects
caused by the fact that the unstable range is finite in the wave-number 
space into the approximate long scale theory. Especially
this is important for the case of small values of $\alpha$,
because in that limit the condition (\ref{local_condition}) for
applicability of the Hamiltonian (\ref{H_l}) becomes too
restrictive. The idea how to improve the approximation is the
following. In general, the exact expression for the Hamiltonian
of a pair of singular filaments, after integration by parts, can
be represented as the half of the integral over a surface $\Sigma$
drawn between the filaments (one half since we consider only one
from two symmetric strings):
$$
{\cal H}_\alpha=\frac{1}{2}\int \frac{({\bf v}\cdot{\bf p})}{2}d{\bf r}=
\frac{\Gamma}{2}\int_{\Sigma}\frac{({\bf v}\cdot d{\bf S})}{2},
$$
because the canonical momentum field ${\bf p}$ created by filaments
is determined by a multi-valued scalar potential $\Phi(\bf r)$:
${\bf p}=\nabla \Phi$, which has the additive increment
$\Gamma=\oint({\bf p}\cdot d{\bf l})$ after
passing around a filament. Also the equality  $\mbox{div }{\bf v}=0$ is
important for derivation of the last expression. In the case of small $\alpha$,
we should just more carefully take account of the contribution to the surface
integral from the vicinity of filaments. As the result of such consideration,
we find that for a better approximation it is sufficient to replace in
(\ref{H_l}) the projection of the arc-length element by the entire arc-length
element and, correspondingly, use the Hamiltonian
\begin{equation}\label{H_l_next}
{\cal H}_l^{\alpha\ll 1}\{{\bf R(\xi)}\}\sim
\oint\frac{Y^\alpha}{\alpha}\sqrt{X'^2+Y'^2+Z'^2}d\xi.
\end{equation}
We stress once more here that this expression is valid only in the case
$\alpha\ll 1$, $Y/L\ll 1$.

\subsection*{Acknowledgments}

These investigations were supported by the Graduate School in
Nonlinear Science (The Danish Research Academy), and by the INTAS. 
The work of V.P.R. was supported by the RFBR (grant No. 00-01-00929) 
and by the Russian State Program of Support of the Leading Scientific 
Schools (grant No. 00-15-96007).

\subsection*{Appendix A}

In order to have some closed expressions for the functions $A_\alpha(k)$ and 
$B_\alpha(k)$ instead of the integral representations (\ref{Ak}) and (\ref{Bk}),
let us use the following mathematical relations \cite{table}:
\begin{equation}
I_{n-\alpha}=\int_0^{+\infty}\!\!\frac{d\zeta}
{(\zeta^2+1)^{\frac{n-\alpha}{2}}}=
\frac{\sqrt{\pi}}{2}\frac{\Gamma(\frac{n-1-\alpha}{2})}
{\Gamma(\frac{n-\alpha}{2})},
\end{equation}
\begin{equation}
I^{(1)} =\! \int_{0}^{+\infty } \!\!\!\cos(kb\zeta)
\frac{d\zeta}{\zeta^{1-\alpha }}=
{{(b k)}^{-\alpha }}\! \cos \left(\frac{\pi\alpha }{2}\right)\Gamma(\alpha),
\end{equation}
\begin{equation}
I^{(3)}\!=\!\int_0^{+\infty}\!\!\!(1\!-\!\cos(kb\zeta))
\frac{d\zeta}{\zeta^{3-\alpha}}
=\frac{(kb)^2 I^{(1)}}{(1-\alpha)(2-\alpha)},
\end{equation}
\begin{equation}\label{table_integral}
\int_{0}^{+\infty}
\frac{\cos(q\zeta) d\zeta} {(\zeta^2+1)^{\rho}}=
\frac{\sqrt{\pi}}{\Gamma(\rho)}
\left(\frac{q}{2}\right)^{\rho-\frac{1}{2}}K_{\rho-\frac{1}{2}}(q),
\quad \rho >0,
\end{equation}
where $\Gamma(x)$ is the Gamma function, $K_\nu(x)$ is the modified Bessel
function of the second kind.
The integral (\ref{table_integral}) results in the equalities
\begin{equation}
J^{(1)}=\int _{0}^{+\infty}\!\!
\frac{\cos(kb\zeta) d\zeta} {(\zeta^2+1)^{\frac{1-\alpha}{2}}}
=\frac{\sqrt{\pi}}{\Gamma\left(\frac{1-\alpha}{2}\right)}
\left(\frac{kb}{2}\right)^{-\frac{\alpha}{2}}K_{-\frac{\alpha}{2}}(kb),
\end{equation}
\begin{equation}
J^{(3)}=\int _{0}^{+\infty}\!\!
\frac{\cos(kb\zeta) d\zeta} {(\zeta^2+1)^{\frac{3-\alpha}{2}}}
=\frac{\sqrt{\pi}}{\Gamma\left(\frac{3-\alpha}{2}\right)}
\left(\frac{kb}{2}\right)^{1-\frac{\alpha}{2}}K_{1-\frac{\alpha}{2}}(kb),
\end{equation}
\begin{equation}
J^{(5)}=\int _{0}^{+\infty}\!\!
\frac{\cos(kb\zeta) d\zeta} {(\zeta^2+1)^{\frac{5-\alpha}{2}}}
=\frac{\sqrt{\pi}}{\Gamma\left(\frac{5-\alpha}{2}\right)}
\left(\frac{kb}{2}\right)^{2-\frac{\alpha}{2}}K_{2-\frac{\alpha}{2}}(kb).
\end{equation}
Thus, we have from (\ref{Ak}) and (\ref{Bk}):
\begin{equation}
A_\alpha(k)=2(1-\alpha)^2 b^{\alpha-2}I^{(3)}
-2k^2b^\alpha J^{(1)}
+2(1-\alpha)b^{\alpha-2}\left(I_{3-\alpha}-J^{(3)}\right),
\end{equation}
\begin{equation}
B_\alpha(k)=2(1-\alpha)^2 b^{\alpha-2}I^{(3)}-2(1-\alpha)(3-\alpha)b^{\alpha-2}
\left(2(J^{(5)}+I_{5-\alpha})-J^{(3)}-I_{3-\alpha}\right).
\end{equation}

\subsection*{Appendix B}

In this Appendix, we explain how the solution
(\ref{varphi_p}-\ref{r_p})
of the system (\ref{r_varphi}-\ref{y_varphi}) can be obtained. Let us
introduce the designations
\begin{equation}
Q=(dr/d\varphi)^2,\qquad s=r^2,
\end{equation}
then consider temporary $s$ as independent variable, and rewrite
 Eq.(\ref{y_varphi}) as follows:
\begin{equation}
y=\frac{2}{\alpha}\left(Q\frac{d}{ds}
\left(\frac{y^\alpha}{\sqrt{s+Q}}\right)+
\frac{y^\alpha(dQ/ds-1)}{2\sqrt{s+Q}}\right),
\end{equation}
or equivalently
\begin{equation}
y=\frac{2}{\alpha}\left(\alpha \frac{dy}{ds}y^{\alpha-1}\sqrt{s+Q}
-\frac{d}{ds}\left(\frac{y^\alpha s}{\sqrt{s+Q}}\right)\right).
\end{equation}
Substituting into this equation the relation
\begin{equation}
Q=s(sy^{2(1-\alpha)}-1)
\end{equation}
which follows from the equation (\ref{r_varphi}),  we have the
following equation for $y(s)$:
\begin{equation}
y=\frac{2}{\alpha}\frac{dy}{ds}
\left(\alpha s+(1-2\alpha)y^{2(\alpha-1)}\right).
\end{equation}
This first order differential equation is linear for the inverse dependence
$s(y)$, and its general solution is
\begin{equation}
s(y)=C^2y^2-\frac{(1-2\alpha)}{\alpha(2-\alpha)}y^{2(\alpha-1)},
\end{equation}
where $C$ is an arbitrary constant of integration.
Thus, we have the relation between $y$ and $s=r^2$. To obtain another relation,
between $y$ and $\varphi$, let us use the equation
\begin{equation}
d\varphi=\frac{ds}{2\sqrt{sQ}},
\end{equation}
which gives us the integral
$$
\varphi-\varphi_0=\int\frac{s'(y)dy}
{2s(y)\sqrt{s(y)y^{2(1-\alpha)}-1}}
=\int\frac{\left(C^2y^{2(2-\alpha)}+
\frac{(1-2\alpha)(1-\alpha)}{\alpha(2-\alpha)}\right)\frac{dy}{y}}
{\left(C^2y^{2(2-\alpha)}-\frac{(1-2\alpha)}{\alpha(2-\alpha)}\right)
\sqrt{C^2y^{2(2-\alpha)}-\frac{(1-\alpha^2)}{\alpha(2-\alpha)}}}=
$$
\begin{equation}
=\arctan\sqrt
{\left(C^2y^{2(2-\alpha)}-\frac{(1-\alpha^2)}{\alpha(2-\alpha)}\right)}
-\sqrt{\frac{\alpha(1-\alpha)}{(2-\alpha)(1+\alpha)}}
\arctan\sqrt{\left(\frac{\alpha(2-\alpha)}{(1-\alpha^2)}C^2y^{2(2-\alpha)}
-1\right)}.
\end{equation}
After introducing the new parameter
\begin{equation}
p=\sqrt{\left(C^2y^{2(2-\alpha)}-
\frac{(1-\alpha^2)}{\alpha(2-\alpha)}\right)},
\end{equation}
we arrive at solution of the system (\ref{r_varphi}-\ref{y_varphi})
in the form (\ref{varphi_p}-\ref{r_p}).

\end{document}